\documentclass[11pt]{article}
\usepackage{mathrsfs}
\usepackage{amsmath}
\usepackage{amsfonts}
\usepackage{amssymb, amsmath, cite}
\usepackage{color}
\usepackage{graphicx}
\usepackage{float}

\newcommand\be{\begin{equation}}
\newcommand\ee{\end{equation}}
\newcommand\ber{\begin{eqnarray}}
\newcommand\eer{\end{eqnarray}}
\newcommand\berr{\begin{eqnarray*}}
\newcommand\eerr{\end{eqnarray*}}\newcommand\eq{\eqref}
\newcommand\bea{\begin{eqnarray}}
\newcommand\eea{\end{eqnarray}}

\newcommand\dd{\mbox{d}}\newcommand\lm{\lambda}
\newcommand\e{\mathrm{e}}
\newcommand\pa{\partial}
\newcommand{\vp}{\varphi}
\newcommand{\nn}{\nonumber}

\title{Dilaton Mass Formulas in a Hairy Binary Black Hole Model}

\author{Shouxin Chen\\School of Mathematics and Statistics\\Henan University\\
Kaifeng, Henan 475004, PR China\\ \\
Yisong Yang\footnote{Corresponding author. Email address: yisongyang@nyu.edu}\\Courant Institute of Mathematical Sciences\\ New York University\\New York, New York 10012, USA
}

\date{}

\begin{document}
\maketitle
\begin{abstract}
In this note an analytic integration  is obtained for the differential equation governing the scalar-field-dependent mass in a hairy binary black hole model,
 in the context of the Einstein--Maxwell--dilation theory, which gives a closed-form formula-level description of the mass function. We also identify a particular solution which attracts
all solutions of the mass-governing equation exponentially rapidly in large-dilaton-field limit.
\medskip

\medskip

\medskip


\end{abstract}

In \cite{GM,GHS}, the Einstein--Maxwell--dilation theory derived from string theory and governed by the action principle
\be\label{1}
S=\int\dd^4\sqrt{-g}\left(R-2\pa_\mu\vp\pa^\mu\vp-\e^{-2a\vp}F_{\mu\nu}F^{\mu\nu}\right),
\ee
is considered, where $R$ is the Ricci scalar induced from the 4-dimensional Minkowski spacetime with metric $g_{\mu\nu}$ of signature $(-+++)$, $F_{\mu\nu}=\pa_\mu A_\nu -\pa_\nu A_\mu$ 
the Maxwell electromagnetic tensor associated with the gauge field $A_\mu$, $\vp$ a real scalar field, or the dilaton field, and $a$ a fundamental coupling parameter which may be assumed to be positive without loss of generality. Using $R_{\mu\nu}$ to denote
the Ricci tensor, $\nabla_\mu$ the covariant derivative
with respect to the spacetime metric, and $\Box=\nabla_\mu\nabla^\mu$ the d'Alembertian, the equations of motion of (\ref{1}) are
\bea
&&\frac12 R_{\mu\nu}=\pa_\mu\vp\pa_\nu\vp+\e^{-2a\vp}\left(F_{\mu\alpha}g^{\alpha\beta}F_{\nu\beta}-\frac14 g_{\mu\nu} F^{\alpha\beta}F_{\alpha\beta}\right),\label{2}\\
&&\nabla_\nu\left(\e^{-2a\vp}F^{\mu\nu}\right)=0,\label{3}\\
&&\Box\vp=-\frac a2\e^{-2a\vp} F_{\mu\nu}F^{\mu\nu}.\label{4}
\eea
It is obtained in \cite{GHS} that the most general static and spherically symmetric solutions which are asymptotically flat and possess regular horizons give rise to a class of electrically charged non-rotating black hole metrics
characterized by three  integration parameters among which two positive ones, $r_+$  and $r_-$, correspond to the outer and inner horizons, respectively, and the third one,  $\vp_0$, represents the background value of the dilaton
field. Specifically, following the presentation in \cite{GHS,HH,J,K}, these free parameters determine the ADM mass $M$,  electric charge $Q$, and dilaton scalar charge $D$ (the ``hair") by the expressions
\bea
M&=&\frac{r_+}2+\frac{r_-}2\left(\frac{1-a^2}{1+a^2}\right),\label{5}\\
Q&=&\sqrt{\frac{r_+ r_-}{1+a^2}}\e^{-a\vp_0},\label{6}\\
D&=&\frac a{1+a^2}r_-,\label{7}
\eea
respectively, such that the solutions to the Einstein--Maxwell--dilaton equations (\ref{2})--(\ref{4}) in terms of the metric tensor $g_{\mu\nu}$, the gauge field $A_\mu$, and the dilaton scalar field $\vp$ are given by
\bea
\dd s^2&=&-A(r)\dd t^2+\frac1{A(r)}\dd r^2+B(r) r^2\dd \Omega^2,\label{8}\\
A_0(r)&=&-\frac Qr\e^{2a\vp_0},\quad A_i(r)=0,\label{9}\\
\vp(r)&=&\vp_0+\frac a{1+a^2}\ln\left(1-\frac{r_-}r\right),\label{10}
\eea
with
\bea
A(r)&=&\left(1-\frac{r_+}r\right)\left(1-\frac{r_-}r\right)^{\frac{1-a^2}{1+a^2}},\label{11}\\
B(r)&=&\left(1-\frac{r_-}r\right)^{\frac{2a^2}{1+a^2}},\label{12}
\eea
with $\dd\Omega^2$ the line element of $S^2$ in terms of the polar and azimuthal angles. From the relations (\ref{5})--(\ref{7}), we get 
\be\label{Deq}
D\left(2M-\frac{1-a^2}a D\right)=aQ^2 \e^{2a\vp_0},
\ee
so that the outer and inner horizon radii may be rewritten as \cite{K}
\bea
r_-&=&\frac{1+a^2}a D\nn\\
&=&\frac{1+a^2}{1-a^2}\left(M\pm\sqrt{M^2-(1-a^2)Q^2\e^{2a\vp_0}}\right),\label{13}\\
r_+&=&2M-\frac{1-a^2}a D\nn\\
&=&M\mp\sqrt{M^2-(1-a^2)Q^2\e^{2a\vp_0}},\label{14}
\eea
which will be used to deduce the mass-governing equation of our interest.

Following \cite{J,K} and the references therein, the action principle of the Einstein--Maxwell-dilaton theory in the presence of a collection of $\vp$-dependent massive and electrically charged point particles,
labeled by $A$, with mass $m_A(\vp)$,  constant charge $q_A$, and 4-coordinates $x^\mu_A$, expands that of (\ref{1}) to read
\be\label{15}
S=\frac1{16\pi}\int\dd^4\sqrt{-g}\left(R-2\pa_\mu\vp\pa^\mu\vp-\e^{-2a\vp}F_{\mu\nu}F^{\mu\nu}\right)-\sum_{A}\int m_A(\vp)\dd s_A+\sum_{A} q_A\int A_\mu\dd x^\mu_A,
\ee
where $\dd s_A=\sqrt{-g_{\mu\nu}\dd x^\mu_A \dd x^\mu_A}$, such that the field equations of motion are updated to be \cite{J}:
\bea
&&\frac12 R_{\mu\nu}=\pa_\mu\vp\pa_\nu\vp+\e^{-2a\vp}\left(F_{\mu\alpha}g^{\alpha\beta}F_{\nu\beta}-\frac14 g_{\mu\nu} F^{\alpha\beta}F_{\alpha\beta}\right)+4\pi\sum_A\left(T^A_{\mu\nu}-\frac12 g_{\mu\nu}T^A\right),\label{16}\\
&&\nabla_\nu\left(\e^{-2a\vp}F^{\mu\nu}\right)=4\pi\sum_A q_A\int \dd s_A \frac{\delta^{(4)}(x-x_A(s_A))}{\sqrt{-g}}\frac{\dd x_A^\mu}{\dd s_A},\label{17}\\
&&\Box\vp=-\frac a2\e^{-2a\vp} F_{\mu\nu}F^{\mu\nu}+4\pi\sum_A \int \dd s_A \frac{\delta^{(4)}(x-x_A(s_A))}{\sqrt{-g}}\frac{\dd m_A}{\dd \vp},\label{18}
\eea
where $T^A=g^{\mu\nu}T^A_{\mu\nu}$ is the contracted energy-momentum tensor.
Thus, the massive and electrically charged point particles generate sources to alter the dynamics of the fields, where $T^A_{\mu\nu}$ is the energy-momentum tensor associated with the $A$th body and given as
\be
T_A^{\mu\nu}(x)=\int \dd s_A  m_A(\vp)\frac{\delta^{(4)}(x-x_A(s_A))}{\sqrt{-g}}\frac{\dd x_A^\mu}{\dd s_A}\frac{\dd x_A^\nu}{\dd s_A},
\ee
which relies on the dilaton field dependent mass $m_A(\vp)$ explicitly. Our main interest here is an analytic determination of the function $m_A(\vp)$.

Following \cite{J,K}, the dilaton field dependent mass, $m_A(\vp)$, may be determined through a matching condition with $M\to m_A(\vp), Q\to q_A, D\to \frac{\dd m_A(\vp)}{\dd\vp}$  in \eqref{Deq}, and replacing the background value of the dilaton scalar $\vp_0$
with its generic value $\vp$,
leading to the following fully nonlinear differential equation \cite{J}:
\be\label{meq}
\left(2m_A(\vp)-\frac{(1-a^2)}a\,\frac{\dd m_A(\vp)}{\dd\vp}\right)\frac{\dd m_A(\vp)}{\dd\vp}=aq^2_A \e^{2a\vp}.
\ee

Note that, due to gravitational radiation and back-reaction, the point-like source formalism here has its limited validity and a 
more comprehensive
consideration will need to involve nonlocal treatments \cite{DB,H,D,C,M,Wa,Po}. See also the review \cite{P} and references therein on this
vast subject.

{\bf Review of known results}

In \cite{J}, a systematic study is carried out for the critical case of the equation \eq{meq}, $a=1$, whose integration renders
the analytic solution
\be\label{ma}
m_A(\vp)=\sqrt{\mu^2_A+q^2_A\frac{\e^{2\vp}}2},
\ee
where $\mu_A>0$ is a massive integration constant. Although this solution is simple, it displays an array of interesting and illustrative
features as insightful hints for the study in the general situation, $a\neq1$. For example, it is seen that a purely massive black hole,
$m_A(\vp)=\mu_A$ occurs exactly in the electrically neutral situation, $q_A=0$,  decoupling itself from the gauge and scalar fields;
through matching conditions at infinity, the result \eq{ma} yields the following formula (with arbitrary $q_A$):
\be
\mu^2_A=\frac14 r_+(r_+-r_-)=\frac{A_H}{16\pi},
\ee
relating the black hole mass to the outer horizon radius $r_+$, inner horizon radius $r_-$, and the area $A_H$ of the horizon
of the black hole. Furthermore, using \eq{ma}, the quantities
\be
\alpha_A(\vp)=\frac{\dd\ln m_A(\vp)}{\dd\vp},\quad \beta_A(\vp)=\frac{\dd\alpha_A(\vp)}{\dd\vp}, 
\ee
regarded fundamental for the computation of the two-body Lagrangian at the post-Keplerian order \cite{J}, may  explicitly be
expressed to give some results of marks, among which is the following:
\be
\alpha_A(\vp)=\frac1{1+\e^{2\left(\ln\left[\frac{\sqrt{2}\mu_A}{|q_A|}\right]-\vp\right)}},
\ee
which is of a Fermi--Dirac distribution function type so that a critical scalar cosmological background level,
\be
\phi_{\mbox{\tiny c}}=\ln\left[\frac{\sqrt{2}\mu_A}{|q_A|}\right],
\ee
is recognized which divides two extreme regimes, decoupled and strongly coupled regimes,
characterized by $\alpha_A\to 0$ and $\alpha_A\to1$, respectively, concerning the coupling between
gravity and scalar and gauge fields. In the general situation when $a$ is arbitrary, it is shown \cite{J} by solving the equation
\eq{meq} numerically that the same Fermi--Dirac distribution function type behavior occurs for the quantity $\alpha_A(\vp)$,
referred to as the scalar charge in \cite{K}, so that a sharp division line exists between the same two extreme regimes,
decoupled and strongly coupled, characterized by $\alpha_A\to0$ and $\alpha_A\to a$, respectively. In this situation, 
it is remarked in \cite{J} that \eq{meq} admits an exact analytic solution in the form of a parametric equation on $m_A(\vp)$
without further elaboration and the solution has the asymptotic form $\frac{|q_A|\e^{a\vp}}{\sqrt{1+a^2}}$ as $\vp\to\infty$.

To proceed further and for convenience, when $a\neq1$, we may resolve the derivative in \eq{meq} by quadrature and suppress the subscript $A$ to arrive  at the equation
\be\label{meq2}
\frac{\dd m(\vp)}{\dd\vp}=\frac a{1-a^2}\left(m(\vp)\pm\sqrt{m^2(\vp)-(1-a^2) q^2\,\e^{2a\vp}}\right),
\ee
which is seen clearly to be non-separable in its present form. Both choices of sign in the equation are of interest.

In \cite{K}, the authors consider (\ref{meq2}) with minus sign, namely, the equation
\be\label{20}
\frac{\dd m(\vp)}{\dd\vp}=\frac a{1-a^2}\left(m(\vp)-\sqrt{m^2(\vp)-(1-a^2)\,q^2 \,\e^{2a\vp}}\right),
\ee
and state that no analytic solution of this equation is known for arbitrary values of $a$ except for the limiting situation $a=1$ (which is understood to be contained in (\ref{meq})). From \eq{20}, the fundamental quantities $\alpha(\vp)$ and $\beta(\vp)$ are obtained \cite{K} to be
\bea
\alpha(\vp)&=&\frac{\dd \ln m(\vp)}{\dd\vp}=\frac{a}{1-a^2}\left(1-\sqrt{1-(1-a^2)\frac{q^2\e^{2a\vp}}{m^2(\vp)}}\right),\label{a}\\
\beta(\vp)&=&\frac{\dd\alpha(\vp)}{\dd\vp}=\frac{a^2 q^2\e^{2a\vp}}{(1-a^2)m^2(\vp)}
\left(1-\frac{a^2}{\sqrt{1-(1-a^2)\frac{q^2\e^{2a\vp}}{m^2(\vp)}}}\right),\label{b}
\eea
whose background-field versions, which are consistent with  \eq{a} and \eq{b}, are deduced in \cite{J} as well by a perturbative argument. Thus, once the mass function $m(\vp)$ is known either analytically or numerically, the properties such as those of
$\alpha(\vp)$ and $\beta(\vp)$ may be obtained correspondingly which enable us to describe the underlying interaction of
gravity and scalar and gauge fields, as in \cite{J,K}. In particular, in \cite{K}, a series of numerical results for $\alpha(\vp)$ 
with $a$ assuming a wide range of values are presented which confirm well the Fermi--Dirac distribution function type behavior
that divides two extreme regimes as investigated in \cite{J}, among other results.
\medskip

Motivated by the results of \cite{J,K} reviewed above, in this note, we aim to obtain a complete analytic integration of the equation \eqref{meq}, or its quadratically resolved version (\ref{meq2}), for arbitrary $a\neq1$.
The equation (\ref{meq2}) is no longer fully nonlinear but it remains non-separable. A key step in our method is to make it separable first.

We begin by considering \eqref{meq2} with negative sign, which is \eqref{20}.

First, we notice that the form of the equation (\ref{20}) mandates the natural constraint
\be\label{Iv}
m^2(\vp)-(1-a^2)q^2 \e^{2a\vp}\geq0
\ee
for the mass function $m(\vp)$, which will be observed in the subsequent discussion. Furthermore, it is easily seen that, for any $a\neq1$, the right-hand side of (\ref{20}) never vanishes. In fact, it always stays positive. In particular,
$m(\vp)$ increases with respect to $\vp$ in all situations.

To proceed, we rewrite \eqref{20} as
\be\label{21}
\frac{\left(m+\e^{a\vp}\sqrt{(m\e^{-a\vp})^2-(1-a^2)q^2}\right)}{\e^{2a\vp}}\,\frac{\dd m}{\dd\vp}=aq^2.
\ee
Now set $u(\vp)=m(\vp)\e^{-a\vp}$ to recast \eqref{21} into
\be\label{22}
\left(u+\sqrt{u^2-(1-a^2)q^2}\right)\left(\frac{\dd u}{\dd \vp}+au\right)=aq^2,
\ee
which can further be put into an equation with separated variables as follows of the form
\be\label{ueq}
\left(u+\sqrt{u^2-\lm^2}\right)\frac{\dd u}{\dd\vp}=a\left(q^2-u\left[u+\sqrt{u^2-\lm^2}\right]\right),
\ee
where $\lm^2=(1-a^2)q^2$. The equation has exactly one equilibrium at
\be\label{eq}
u=\frac q{\sqrt{1+a^2}},
\ee
giving rise to an analytic solution for the mass function:
\be\label{exp}
m(\vp)=\frac {q\e^{a\vp}}{\sqrt{1+a^2}}.
\ee
(Negative equilibrium is irrelevant for our problem and thus discarded.) It is readily examined that the solution \eqref{exp} fulfills the constraint (\ref{Iv}). (Note that this solution may also be deduced
with the ansatz $m(\vp)=C\e^{a\vp}$ in (\ref{meq}) for arbitrary $a$ including $a=1$.)

In order to get other solutions of the equation (\ref{ueq}), we consider the initial-value condition
\be\label{Ic}
u(\vp_0)=u_0\neq \frac q{\sqrt{1+a^2}}\quad \mbox{or}\quad m(\vp_0)=m_0\neq \frac {q\e^{a\vp_0}}{\sqrt{1+a^2}}.
\ee
That is, we consider non-equilibrium solutions of (\ref{ueq}). Thus, the right-hand side of (\ref{ueq}) is nonzero which allows us to recast the equation into
\be\label{23}
\frac{u+\sqrt{u^2-\lm^2}}{q^2-u\left(u+\sqrt{u^2-\lm^2}\right)}\,\dd u=a\dd\vp,
\ee

(i) Assume $a<1$. Then $\lm^2>0$. So in view of (\ref{Iv}), we have 
\be
u(\vp)=m(\vp)\e^{-a\vp}\geq \lm=q\sqrt{1-a^2}.
\ee
 Thus we may introduce the new variable
\be\label{conv}
0\leq w=\sqrt{\frac{u-\lm}{u+\lm}}<1 \quad\mbox{with the conversion}\quad u=\frac{\lm(w^2+1)}{1-w^2}.
\ee
Consequently the integration of the left-hand side of \eqref{23} reads
\bea\label{25}
I&=&\int\frac{u+\sqrt{u^2-\lm^2}}{q^2-u\left(u+\sqrt{u^2-\lm^2}\right)}\,\dd u\nn\\
&=&-\frac{4\lm^2}{q^2}\int\frac w{(w^2-1)(a^2w^2-2w+a^2)}\,\dd w.
\eea
Write $w_{\pm}=\frac1{a^2}\left(1\pm\sqrt{1-a^4}\right)$. Then $0<w_-<w_+$ which leads to the following standard partial-fractional decomposition 
\bea\label{26}
\frac w{(w^2-1)(a^2 w^2-2w+a^2)}&=&-\frac 1{4(1-a^2)(w-1)}+\frac1{4(1+a^2)(w+1)}\nn\\
&&+\frac{a^2}{4(1-a^4)}\left(\frac1{w-w_-}+\frac1{w-w_+}\right).
\eea
It is worth to note that $w_+>1$ and $w_-$ corresponds to the equilibrium (\ref{eq}). That is, we have
\be
0<w=\sqrt{\frac{u-\lm}{u+\lm}}=\frac1{a^2}\left(1-\sqrt{1-a^4}\right)=w_-<1\quad\mbox{when }u=\frac q{\sqrt{1+a^2}}.
\ee
In other words,  $w=w_-$, $0<w_-<1$, is the only singular point in the integrand of \eqref{25} which should be attended.

Hence, inserting (\ref{26}) into (\ref{25}), and with the observation (\ref{Ic}) such that
\be
 w_0=w(\vp_0)=\sqrt{\frac{m(\vp_0)\e^{-a\vp_0}-\lm}{m(\vp_0)\e^{-a\vp_0}+\lm}}=\sqrt{\frac{m(\vp_0)\e^{-a\vp_0}-q\sqrt{1-a^2}}{m(\vp_0)\e^{-a\vp_0}+q\sqrt{1-a^2}}}\neq w_-,
\ee
we obtain the integration of (\ref{ueq})  to be
\be\label{wsol}
\ln|w(\vp)-1|-\left(\frac{1-a^2}{1+a^2}\right)\ln|w(\vp)+1|-\frac{a^2}{1+a^2}\ln|a^2 w^2(\vp)-2w(\vp)+a^2|=C_0+a(\vp-\vp_0),
\ee
where
\bea
C_0&=&\ln|w_0-1|-\left(\frac{1-a^2}{1+a^2}\right)\ln|w_0+1|-\frac{a^2}{1+a^2}\ln|a^2 w_0^2-2w_0+a^2|,\\
w(\vp)&=&\sqrt{\frac{m(\vp)-\sqrt{1-a^2}\,q\,\e^{a\vp}}{m(\vp)+\sqrt{1-a^2}\,q\,\e^{a\vp}}}.
\eea
In particular, we obtain the asymptotic behavior $w(\vp)\to w_-$ as $\vp\to\infty$, leading to 
\be\label{asym}
\lim_{\vp\to\infty} u(\vp)=\frac q{\sqrt{1+a^2}}.
\ee
So, as a by-product, the analytic solution (\ref{exp}) is identified as the  large dilaton field limit of all solutions to the mass governing equation \eqref{20} when $a<1$.

It may be useful to obtain some estimate for the convergence result (\ref{asym}). In fact, from (\ref{wsol}), we can get for $\vp$ large the asymptotic estimate
\be\label{w1}
|w(\vp)-w_-|=\mbox{O}\left(\e^{-\left(a+\frac1a\right)\vp}\right).
\ee
Inserting this result back to $u(\vp)$, we obtain
\be\label{u1}
\left|u(\vp)-\frac q{\sqrt{1+a^2}}\right|=\mbox{O}\left(\e^{-\left(a+\frac1a\right)\vp}\right).
\ee
Therefore, applying (\ref{u1}) to the relation $m(\vp)=u(\vp)\e^{a\vp}$, we find
\be\label{m1}
m(\vp)=\frac{q\e^{a\vp}}{\sqrt{1+a^2}}+\mbox{O} \left(\e^{-\frac\vp a}\right),\quad\vp\to\infty.
\ee

(ii) Assume $a>1$. Set $\sigma^2=(a^2-1)q^2=-\lm^2$ and rewrite the integration of the left-hand side of (\ref{23}) as
\bea\label{Iw}
I&=&\int\frac{u+\sqrt{u^2+\sigma^2}}{q^2-u\left(u+\sqrt{u^2+\sigma^2}\right)}\,\dd u\nn\\
&=&-\frac{\sigma^2}{(a^2+1)q^2}\int\frac{\sigma^2 +w^2}{w(w^2-\kappa^2)}\,\dd w,
\eea
where we have set
\be\label{wu}
w=\sqrt{u^2+\sigma^2}-u\quad\mbox{or}\quad u=\frac{\sigma^2 -w^2}{2w},\quad \kappa^2=\sigma^2\left(\frac{a^2-1}{a^2+1}\right),\quad \kappa>0.
\ee
Note that the only singular point in the integrand of \eqref{Iw} involving the variable $w$ occurs at $w=\kappa$
which corresponds to \eqref{eq} again. Thus, with 
\be\label{ii47}
w(\vp_0)=w_0\neq \kappa\quad \mbox{or}\quad u(\vp_0)=u_0\neq\frac{q}{\sqrt{a^2+1}},
\ee
we arrive at the solution
\be\label{ii}
\frac{a^2}{a^2+1}\ln\left|w^2(\vp)-\frac{(a^2-1)^2 q^2}{a^2+1}\right|-\ln w(\vp)=C_0- a(\vp-\vp_0),
\ee
where
\bea
C_0&=&\frac{a^2}{a^2+1}\ln\left|w_0^2-\frac{(a^2-1)^2 q^2}{a^2+1}\right|-\ln w_0,\quad w_0\neq \frac{(a^2-1) q}{\sqrt{a^2+1}}=\kappa,
\label{ii0}\\
w(\vp)&=& \sqrt{(m(\vp)\,\e^{-a\vp})^2+(a^2-1)q^2}-m(\vp)\,\e^{-a\vp}.\label{iim}
\eea
Hence, again, $w(\vp)\to\kappa$ as $\vp\to\infty$. That is, the same asymptotic behavior (\ref{asym}) holds as in the situation $a<1$. Moreover, using (\ref{ii}), we may readily deduce the
same precise asymptotic estimates (\ref{w1})--(\ref{m1}) for our solution here.

\begin{figure}[h]
\begin{center}
\includegraphics[height=6cm,width=8cm]{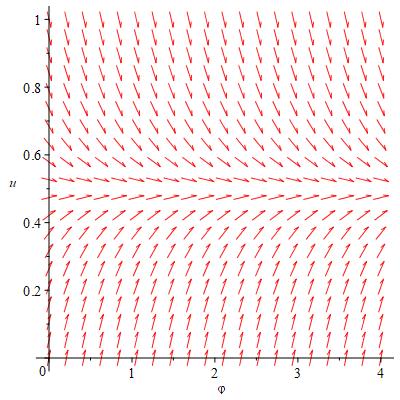}
\caption{A plot of the slope field distribution for the reduced mass equation \eqref{ueq}. It is clearly seen that all solutions are attracted to the unique equilibrium}
\label{F1}
\end{center}
\end{figure}

As an illustration, we present in Figure \ref{F1} a plot of the slope field distribution of the reduced mass-function equation \eqref{ueq} with the data $q=1, a=\sqrt{3}$ such that the equilibrium \eqref{eq} 
is $u=\frac12$. The field is depicted over the region $0<\vp<4, 0<u<1$. All ``trajectories" converge to the horizontal centerline monotonically. 

We now consider (\ref{meq2}) with plus sign:
\be\label{20+}
\frac{\dd m(\vp)}{\dd\vp}=\frac a{1-a^2}\left(m(\vp)+\sqrt{m^2(\vp)-(1-a^2)\,q^2 \,\e^{2a\vp}}\right).
\ee
As before, we may set $u(\vp)=m(\vp)\e^{-a\vp}$ and $\lm^2=(1-a^2)q^2$ to transform \eqref{20+} into a separable equation,
\be\label{ueq+}
\left(u-\sqrt{u^2-\lm^2}\right)\frac{\dd u}{\dd\vp}=a\left(q^2-u\left[u-\sqrt{u^2-\lm^2}\right]\right).
\ee
It is interesting to note that, unlike before, this equation has no positive equilibrium.

 Integrating (\ref{ueq+}), we have
\be\label{I+}
I=\int \frac{\left(u-\sqrt{u^2-\lm^2}\right)}{\left(q^2-u\left[u-\sqrt{u^2-\lm^2}\right]\right)}\,\dd u=a\vp.
\ee

 We again consider two cases.

(iii) If $a<1$, then we use
\be\label{conv+}
0\leq w=\sqrt{\frac{u-\lm}{u+\lm}}<1 \quad\mbox{with the conversion}\quad u=\frac{\lm(w^2+1)}{1-w^2},
\ee
to recast the left-hand side of \eqref{I+} into
\be\label{II+}
I=-\frac{4\lm^2}{q^2}\int\frac w{(w^2-1)(a^2 w^2 +2w+a^2)}\,\dd w.
\ee
The second factor in the denominator of the integrand has roots $w_{\pm}=\frac1{a^2}\left(-1\pm\sqrt{1-a^4}\right)$, both being negative and hence of no concern for the interest of our problem.
Thus, for any initial value
\be
 w_0=w(\vp_0)=\sqrt{\frac{m(\vp_0)\e^{-a\vp_0}-\lm}{m(\vp_0)\e^{-a\vp_0}+\lm}}=\sqrt{\frac{m(\vp_0)\e^{-a\vp_0}-q\sqrt{1-a^2}}{m(\vp_0)\e^{-a\vp_0}+q\sqrt{1-a^2}}},
\ee
we can carry out the integration \eqref{II+} to get the solution to be
\be\label{wsol+}
\left(\frac{1-a^2}{1+a^2}\right)\ln|w(\vp)-1|-\ln|w(\vp)+1|+\frac{a^2}{1+a^2}\ln|a^2 w^2(\vp)+2w(\vp)+a^2|=C_0-a(\vp-\vp_0),
\ee
where
\bea
C_0&=&\left(\frac{1-a^2}{1+a^2}\right)\ln|w_0-1|-\ln|w_0+1|+\frac{a^2}{1+a^2}\ln|a^2 w_0^2+2w_0+a^2|,\\
w(\vp)&=&\sqrt{\frac{m(\vp)-\sqrt{1-a^2}\,q\,\e^{a\vp}}{m(\vp)+\sqrt{1-a^2}\,q\,\e^{a\vp}}}.
\eea

(iv) If $a>1$, we set $\sigma^2=(a^2-1)q^2=-\lm^2$ ($\sigma>0$) and 
\be\label{conv++}
\sigma<w=u+\sqrt{u^2+\sigma^2}\quad\mbox{with the conversion}\quad u=\frac{w^2-\sigma^2}{2w},
\ee
to transform the integral $I$ in (\ref{I+}) into
\be\label{Iw+}
I=-\frac{\sigma^2}{q^2}\int\frac{\left(w^2+\sigma^2\right)}{w\left((1+a^2)w^2-(a^2-1) \sigma^2\right)}\,\dd w,
\ee
which is the same as \eqref{Iw}. Hence we arrive at the solution (\ref{ii}) with arbitrary initial condition (\ref{ii0}) (without the restriction (\ref{ii47}) since (\ref{ueq+}) has no equilibrium for the interest of our problem) and
\be
w(\vp)=\sqrt{(m(\vp)\,\e^{-a\vp})^2+(a^2-1)q^2}+m(\vp)\,\e^{-a\vp}.
\ee

In conclusion, for the dilaton mass equation (\ref{meq2}) with the minus sign, i.e., the equation (\ref{20}),  for any $a>0, a\neq1$, the mass function $m(\vp)$ governed by the equation is given by the formulas 
\bea
&&\ln\left|\sqrt{\frac{m(\vp)-\sqrt{1-a^2}\,q\,\e^{a\vp}}{m(\vp)+\sqrt{1-a^2}\,q\,\e^{a\vp}}}-1\right|-\left(\frac{1-a^2}{1+a^2}\right)\ln\left|\sqrt{\frac{m(\vp)-\sqrt{1-a^2}\,q\,\e^{a\vp}}{m(\vp)+\sqrt{1-a^2}\,q\,\e^{a\vp}}}+1\right|\nn\\
&&-\frac{a^2}{1+a^2}\ln\left|\frac{m(\vp)-\sqrt{1-a^2}\,q\,\e^{a\vp}}{m(\vp)+\sqrt{1-a^2}\,q\,\e^{a\vp}}-\frac2{a^2}\sqrt{\frac{m(\vp)-\sqrt{1-a^2}\,q\,\e^{a\vp}}{m(\vp)+\sqrt{1-a^2}\,q\,\e^{a\vp}}}+1\right|\nn\\
&&=C_0+a(\vp-\vp_0),\label{mf1}
\eea
where
\bea\label{mf2}
C_0 &=&\ln\left|\sqrt{\frac{m_0-\sqrt{1-a^2}\,q\,\e^{a\vp_0}}{m_0+\sqrt{1-a^2}\,q\,\e^{a\vp_0}}}-1\right|-\left(\frac{1-a^2}{1+a^2}\right)\ln\left|\sqrt{\frac{m_0-\sqrt{1-a^2}\,q\,\e^{a\vp_0}}{m_0+\sqrt{1-a^2}\,q\,\e^{a\vp_0}}}+1\right|\nn\\
&&-\frac{a^2}{1+a^2}\ln\left|\frac{m_0-\sqrt{1-a^2}\,q\,\e^{a\vp_0}}{m_0+\sqrt{1-a^2}\,q\,\e^{a\vp_0}}-\frac2{a^2}\sqrt{\frac{m_0-\sqrt{1-a^2}\,q\,\e^{a\vp_0}}{m_0+\sqrt{1-a^2}\,q\,\e^{a\vp_0}}}+1\right|,
\eea
when $a<1$, and 
\bea\label{mf3}
&&\frac{a^2}{a^2+1}\ln\left|\left( \sqrt{m^2(\vp)+(a^2-1)q^2\,\e^{2a\vp}}-m(\vp)\right)^2-\frac{(a^2-1)^2 q^2\,\e^{2a\vp}}{a^2+1}\right|\nn\\
&&-\ln \left( \sqrt{m^2(\vp)+(a^2-1)q^2\,\e^{2a\vp}}-m(\vp)\right)
\nn\\
&&=C_0-\frac{2 a}{a^2+1}(\vp-\vp_0),
\eea
where
\bea\label{mf4}
C_0&=&\frac{a^2}{a^2+1}\ln\left|\left( \sqrt{m^2_0+(a^2-1)q^2\,\e^{2a\vp_0}}-m_0\right)^2-\frac{(a^2-1)^2 q^2\,\e^{2a\vp_0}}{a^2+1}\right|\nn\\
&&-\ln \left( \sqrt{m^2_0+(a^2-1)q^2\,\e^{2a\vp_0}}-m_0\right),
\eea
when $a>1$, both with $m_0=m(\vp_0)\neq \frac{q\e^{a\vp_0}}{\sqrt{1+a^2}}$. Moreover,  the mass function $m(\vp)$ in both cases assumes its asymptotic form
\be
m(\vp)\sim \frac{q\e^{a\vp}}{\sqrt{1+a^2}},\quad \vp\gg1,
\ee
exponentially rapidly, in the sense that
\be
m(\vp)=\frac{q\e^{a\vp}}{\sqrt{1+a^2}}+\mbox{O}\left(\e^{-\frac{\vp}a}\right),\quad\mbox{as }\vp\to\infty.
\ee

 On the other hand, for the dilaton mass equation (\ref{meq2}) with the plus sign, i.e., the equation (\ref{20+}),  the situation is much different. In this situation, for any $a>0, a\neq1$, the transformed equation (\ref{ueq+}) has no equilibrium, or equivalently,
(\ref{20+}) has no particular solution of the form $C\e^{a\vp}$, and 
the mass function $m(\vp)$ governed by the equation is given by the formulas 
\bea\label{mf5}
&&\ln\left|\sqrt{\frac{m(\vp)-\sqrt{1-a^2}\,q\,\e^{a\vp}}{m(\vp)+\sqrt{1-a^2}\,q\,\e^{a\vp}}}+1\right|-\left(\frac{1-a^2}{1+a^2}\right)\ln\left|\sqrt{\frac{m(\vp)-\sqrt{1-a^2}\,q\,\e^{a\vp}}{m(\vp)+\sqrt{1-a^2}\,q\,\e^{a\vp}}}-1\right|\nn\\
&&-\frac{a^2}{1+a^2}\ln\left|\frac{m(\vp)-\sqrt{1-a^2}\,q\,\e^{a\vp}}{m(\vp)+\sqrt{1-a^2}\,q\,\e^{a\vp}}+\frac2{a^2}\sqrt{\frac{m(\vp)-\sqrt{1-a^2}\,q\,\e^{a\vp}}{m(\vp)+\sqrt{1-a^2}\,q\,\e^{a\vp}}}+1\right|\nn\\
&&=C_0+a(\vp-\vp_0),
\eea
where
\bea\label{mf6}
C_0 &=&\ln\left|\sqrt{\frac{m_0-\sqrt{1-a^2}\,q\,\e^{a\vp_0}}{m_0+\sqrt{1-a^2}\,q\,\e^{a\vp_0}}}+1\right|-\left(\frac{1-a^2}{1+a^2}\right)\ln\left|\sqrt{\frac{m_0-\sqrt{1-a^2}\,q\,\e^{a\vp_0}}{m_0+\sqrt{1-a^2}\,q\,\e^{a\vp_0}}}-1\right|\nn\\
&&-\frac{a^2}{1+a^2}\ln\left|\frac{m_0-\sqrt{1-a^2}\,q\,\e^{a\vp_0}}{m_0+\sqrt{1-a^2}\,q\,\e^{a\vp_0}}+\frac2{a^2}\sqrt{\frac{m_0-\sqrt{1-a^2}\,q\,\e^{a\vp_0}}{m_0+\sqrt{1-a^2}\,q\,\e^{a\vp_0}}}+1\right|,
\eea
when $a<1$, and 
\bea\label{mf7}
&&\frac{a^2}{a^2+1}\ln\left|\left( \sqrt{m^2(\vp)+(a^2-1)q^2\,\e^{2a\vp}}+m(\vp)\right)^2-\frac{(a^2-1)^2 q^2\,\e^{2a\vp}}{a^2+1}\right|\nn\\
&&-\ln \left( \sqrt{m^2(\vp)+(a^2-1)q^2\,\e^{2a\vp}}+m(\vp)\right)
\nn\\
&&=C_0-\frac{2 a}{a^2+1}(\vp-\vp_0),
\eea
where
\bea\label{mf8}
C_0
&=&\frac{a^2}{a^2+1}\ln\left|\left( \sqrt{m^2_0+(a^2-1)q^2\,\e^{2a\vp_0}}+m_0\right)^2-\frac{(a^2-1)^2 q^2\,\e^{2a\vp_0}}{a^2+1}\right|\nn\\
&&-\ln \left( \sqrt{m^2_0+(a^2-1)q^2\,\e^{2a\vp_0}}+m_0\right),
\eea
when $a>1$. In this situation, the initial value condition has no restriction and may be arbitrarily prescribed.

It will be useful to show how the analytic solutions obtained in this note help with gaining a precise knowledge on the mass function despite the complexity of the formulas 
(\ref{mf1})--(\ref{mf4}) and (\ref{mf5})--(\ref{mf8}) given in various cases. Indeed, since these formulas are implicit in the mass function $m(\vp)$, it is not a transparent task to
describe how $m(\vp)$ depends on $\vp$. Nevertheless, since in all cases the intermediate variable $w$ depends on $\vp$ monotonically, we may acquire the dependence knowledge of the mass function
on the dilaton field $\vp$ through 
its dependence relation on $w$ relatively much more easily and straightforwardly, as we now demonstrate.

To illustrate, we consider the richer minus-sign equation (\ref{20}), which reduces to (\ref{ueq}), and  choose the data $\vp_0=0, q=1, a=\sqrt{3}$, such that the equilibrium solution for \eqref{ueq}  is $u=\frac12$.
That is, we need to consider solutions satisfying $m_0=m(0)\neq \frac12$ in order to stay away from the resulting particular solution (i.e., the equilibrium). In our approach here, we use (\ref{ii}) to determine the range of $w$ between its
initial state and its asymptotic state as $\vp\to\infty$, which happens to be 1,  and then apply (\ref{wu}) to obtain how $m$ depends on $w$ directly, which is given by the {\em explicit} formula
\be
m(w)=\left(\frac{\sigma^2 -w^2}{2w}\right)\e^{a\vp(w)}.
\ee

\begin{figure}[H]
\begin{center}
\includegraphics[height=6cm,width=7cm]{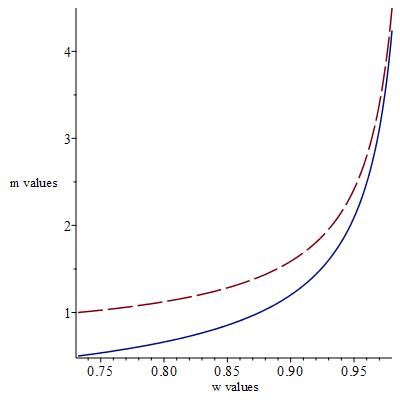}
\caption{A computer-generated plot showing the dependence of the mass function starting from $m_0=1$ on the dilaton field through an intermediate variable $w$ compared with the particular exponential solution. The solution in long-dash curve
approaches the particular solution in solid curve exponentially rapidly in the large $\vp$ limit such that $w\to1$ from the left. }
\label{F2}
\end{center}
\end{figure}

We first present the solution with $m=1$ when $\vp=0$. In this situation, the initial state of $w$ is $\sqrt{3}-1$ in view of (\ref{wu})  since $u=1$ when $\vp=0$. Thus we obtain the range of $w$ to be
\be
\sqrt{3}-1<w<1.
\ee 
In Figure \ref{F2}, we plot the solution represented with the long-dash curve and compare it with the particular solution $m(w)=\frac12 \e^{a\vp(w)}$ represented with solid curve. Initially these two solutions are far apart. However, as $w$ tends to its
limiting value $w=1$ (as $\vp\to\infty$) from below, the two curves stay arbitrarily close to each other rather rapidly.

\begin{figure}[H]
\begin{center}
\includegraphics[height=6cm,width=7cm]{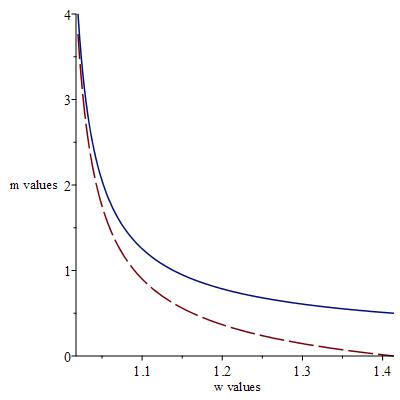}
\caption{A computer-generated plot showing the dependence of the mass function starting from $m_0=0$ on the dilaton field through an intermediate variable $w$ compared with the particular exponential solution. The solution in long-dash curve
approaches the particular solution in solid curve exponentially rapidly in the large $\vp$ limit such that $w\to1$ from the right. }
\label{F3}
\end{center}
\end{figure}

We next present the solution with $m=0$ when $\vp=0$. In this situation, the initial state of $w$ is $\sqrt{2}$ in view of (\ref{wu})  since $u=0$ when $\vp=0$. So we obtain the range of $w$ to be
\be
1<w<\sqrt{2}.
\ee 
In Figure \ref{F3}, we plot the solution represented with the long-dash curve and compare it again with the particular solution $m(w)=\frac12 \e^{a\vp(w)}$ represented with solid curve. These two solutions are  initially far apart. As $w$ tends to its
limiting value $w=1$ (as $\vp\to\infty$) from above, the two curves converge to each other exponentially fast as before.

These examples are concrete realizations of the asymptotic result (\ref{m1}) in the situation of the choice of the minus sign in \eqref{meq2}.
\medskip

Finally, we study the situation of the choice of the plus sign in (\ref{meq2}). For simplicity, we consider the solution discussed in (iv) with $a>1$. Now, with the variable $w$ given in (\ref{conv++}), the equation (\ref{ueq+}) becomes
\be\label{76}
\frac{\dd w}{\dd\vp}=-\frac a{(a^2-1)}\frac{w\left((a^2+1)w^2-(a^2-1)^2 q^2\right)}{\left(w^2+(a^2-1)q^2\right)}.
\ee
From $w\geq\sigma=\sqrt{a^2-1}\,q$ for $u\geq0$, we have 
\be
(a^2+1)w^2-(a^2-1)^2 q^2\geq 2(a^2-1)q^2,
\ee
 leading to $\frac{\dd w}{\dd\vp}<0$ by virtue of  \eqref{76}, which may then be strengthened into
\be
\frac{\dd w}{\dd\vp}\leq -\frac{2a\sqrt{a^2-1}\,q^3}{\left(w_0^2+(a^2-1)q^2\right)},\quad w_0=w(\vp_0).
\ee
This bound implies that there is a finite ``later" stage, $\vp_1>\vp_0$, at which $w(\vp_1)=\sigma$, where $w(\vp)$ solves (\ref{76}) with $w(\vp_0)=w_0>\sigma$ or
$u(\vp_0)>0$.
Beyond $\vp_1$, $w(\vp)<\sigma$ such that $u(\vp)<0$ so that $u(\vp)$ ceases to be meaningful for our problem since $u(\vp)$ is related to the dilaton mass by the expression $u(\vp)=m(\vp)\e^{-a\vp}$,
which is supposed to assume non-negative values.
Therefore, in the present circumstance, the range of $w$ is identified to be
\be\label{79}
\sqrt{a^2-1} \, q=\sigma \leq w\leq w_0.
\ee
Thus we may use (\ref{ii})--(\ref{ii0}) over the range  \eqref{79} to gain knowledge on $\vp$ and then \eqref{conv++} to obtain the mass function as
\be\label{80}
m(w)=\left(\frac{w^2-\sigma^2}{2w}\right)\e^{a\vp(w)}.
\ee

\begin{figure}[H]
\begin{center}
\includegraphics[height=6cm,width=7cm]{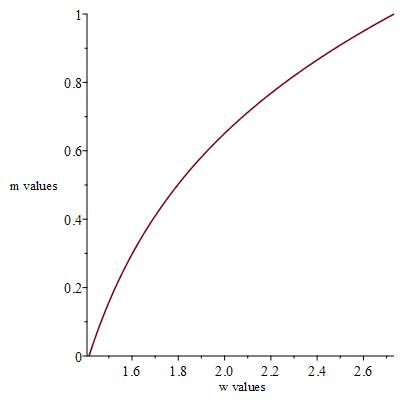}
\caption{In this example, the dilaton mass assumes a positive initial value and eventually decreases to zero at a finite later value of the dilaton field $\vp$. The figure shows that, with $a=\sqrt{3}$ and $q=1$, as $\vp$ increases, the intermediate variable $w$
decreases from its initial value $1+\sqrt{3}$ to its terminal value $\sigma=\sqrt{2}$, which forces $m$ vanish then.  }
\label{F4}
\end{center}
\end{figure}

With the above preparation, we are ready to present an example as another illustration. We choose again $\vp_0=0, q=1, a=\sqrt{3}$. Hence $\sigma=\sqrt{2}$. Set $m_0=1$. Then, in correspondence, we have $w_0=1+\sqrt{3}$.
Inserting these data into (\ref{79}) and (\ref{80}), we arrive at the solution shown in Figure \ref{F4} which shows that as $\vp$ increases the dilaton mass starting from a positive value decreases gradually to zero.

\medskip

{\bf Comparison of results and comments}

To conclude, we discuss some concrete examples and show how to apply the explicit solutions 
obtained in this note to consolidate our knowledge about
the role played by the scalar-field dependent mass function $m(\vp)$ in the unified theory. To be specific, we focus our attention on the scalar charge function
$\alpha(\vp)$ which divides two extreme regimes as noted in \cite{J}. To be in line with \cite{J,K}, we now consider \eq{20}.

First, we note that the newly found special analytic solution \eq{exp}, valid for any value of $a$, is not contained in 
the class of solutions with $\mu_A=m(-\infty)>0$.
 In fact, \eq{exp} belongs to the zero-mass solution $\mu_A=m(-\infty)=0$. For $a=1$, \eq{exp} is contained in \eq{ma} with setting
$\mu_A=0$.
Furthermore, inserting \eq{exp} into
\eq{a}, we have
\be
\alpha(\vp)\equiv a,
\ee
which indicates that the distinguished Fermi--Dirac distribution function type behavior of $\alpha(\vp)$ at $\mu_A>0$ breaks down
for any $a$ at this
exceptional situation (when $\mu_A=0$).

Next, consider $a\neq1$.
As noted in \cite{J}, we have $m(\vp)\to \mu_A$ as $\vp\to-\infty$ with prescribed $\mu_A>0$. So it is straightforward from
\eq{a} that $\alpha(\vp)\to0$ as $\vp\to-\infty$ and we arrive at the decoupled regime as shown in \cite{J}.
On the other hand, however, the structure of \eq{a} indicates that more caution is needed when studying 
the limit of $\alpha(\vp)$ as
$\vp\to\infty$ due to the exponential growth factor, $\e^{2a\vp}$, present in \eq{a}. Fortunately, we have \eq{m1}. Thus, inserting 
\eq{m1} into \eq{a}, we have
\be\label{xxa}
\alpha(\vp)\to a,\quad \vp\to\infty,
\ee
which confirms the result in \cite{J} obtained based on numerical solutions, used to identify another extreme regime where
gravity and  scalar and gauge fields are
strongly coupled, also called the scalarized regime \cite{J}.

\begin{figure}[H]
\begin{center}
\includegraphics[height=6cm,width=7cm]{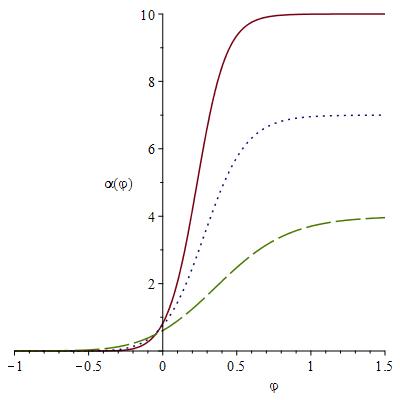}
\caption{The plots of the profiles of the scalar charge function $\alpha(\vp)$ with $q=1$ and $a=4,7,10$. The behavior that
$a(\vp)\to0$ as $\vp\to-\infty$, and $\alpha(\vp)\to a=4,7,10$, respectively, as $\vp\to\infty$, is clearly exhibited.  }
\label{F5}
\end{center}
\end{figure}

Finally we present some results obtained from our exact solutions of the scalar mass equation and compare the results with those 
in \cite{J,K} based on numerical solutions. For this we choose $q=1$ and $a>1$ in \eq{20} and use the solution given by 
\eq{ii}--\eq{iim}. Although these expressions are complicated, we may simplify our computation by rewriting the scalar charge
function $\alpha(\vp)$ given in \eq{a} into the   form
\be
\alpha(\vp)=\frac{a}{1-a^2}\left(1-\sqrt{1-(1-a^2)\left[\frac{\e^{a\vp}}{m(\vp)}\right]^2}\right).\label{aa}
\ee
This is convenient to use because \eq{iim} yields the relation
\be\label{iime}
\frac{m(\vp)}{\e^{a\vp}}=\frac{a^2-1-w^2(\vp)}{2w(\vp)},
\ee
where $w(\vp)$ is determined by \eq{ii} such that the interval $-\infty<\vp<\infty$ corresponds to the interval
\be
0<w<\kappa=\frac{a^2-1}{\sqrt{a^2+1}}.
\ee
In other words, if we use $w$ as an intermediate variable, the scalar charge function $\alpha(\vp)$ is explicitly given by
\be
\alpha(\vp(w))=\frac{a}{1-a^2}\left(1-\sqrt{1-(1-a^2)\left[\frac{2w}{a^2-1-w^2}\right]^2}\right),\quad 0<w<\kappa=\frac{a^2-1}{\sqrt{a^2+1}}, \label{aaw}
\ee
by virtue of \eq{aa} and \eq{iime}. Hence the behavior $\alpha(\vp)\to0$ as $\vp\to-\infty$ and $\alpha(\vp)\to a$ as $\vp\to\infty$ are easily seen in the limits $w\to0$ and $w\to\kappa$, respectively, directly, as well as indirectly as $\vp\to-\infty$ and
$\vp\to\infty$, respectively, via \eq{ii}. In Figure \ref{F5}, we present the plots of $\alpha(\vp)$ obtained this way
for $a=4,7,10$. 

The results in Figure \ref{F5} demonstrate that sharper transitions between decoupled and strongly coupled
gravity and scalar-gauge field regimes are achieved by larger values of
the dilaton coupling parameter $a$ and that the manner of
the transitions is universally of the Fermi--Dirac distribution function type. These results confirm well those obtained in \cite{J,K}.

\medskip

{The authors would like to thank an anonymous referee whose suggestions helped improve the presentation of the paper.}

\medskip

{On behalf of all authors, the corresponding author states that there is no conflict of interest. }

 \end{document}